\documentstyle[12pt]{article}
\input{psfig}
\newcommand{\gsim}
{\mbox{ \begin{picture}(12,12)\put(0,-3){$\sim$}\put(0,2){$>$}
	\end{picture}} }
\newcommand{\lsim}
{\mbox{ \begin{picture}(12,12)\put(0,-3){$\sim$}\put(0,2){$<$}
	\end{picture}} }
\begin{document}
\baselineskip 25pt
\vskip1.0cm
\begin{center}
\Large 	Maximally-Mixed Three Generations of Neutrinos and
	the Solar and Atmospheric Neutrino Problems
\end{center}
\vskip1.5cm
\centerline{\large C. W. Kim and J. A. Lee }
\centerline{\it Department of Physics and Astronomy }
\vskip -0.25truecm
\centerline{\it The Johns Hopkins University }
\vskip -0.25truecm
\centerline{\it Baltimore, Maryland 21218, USA }
\vskip2.0truecm

\centerline{\large Abstract }
\vspace*{0.3truecm}
Motivated by the indication that both the Solar and the
atmospheric neutrino puzzles may simultaneously be solved by
(vacuum as well as matter-induced
resonant) oscillations of two generations of neutrinos with large mixing,
we have analyzed the data on the Solar and atmospheric neutrinos
assuming that all {\it three} neutrinos are maximally mixed.
It is shown that the results of two-generation analyses
are still valid even in the three-generation scheme, {\it i.e.},
the two puzzles can be solved simultaneously if
	$\Delta m_{21}^2 = m_2^2-m_1^2\simeq 10^{-10}\mbox{eV}^2$
and
	$\Delta m_{31}^2 = 10^{-3}\sim 10^{-1}\mbox{eV}^2$.
We have also demonstrated explicitly that with the use of the
see-saw mechanism
it is possible to have large or maximal mixings for neutrinos
even though their masses are highly non-degenerate.

\newpage
\section{Introduction}

It has long been known that the measured Solar neutrino event rates by
three types of experimental detectors
\cite{gallex92,abazov91,lande90,hirata91}
all have shown
the deficiency compared with the Standard Solar Model (SSM) predictions
by
Bahcall and others \cite{bahcallpinsonneault92,turckchiez88}.
The latest ratios of the observed event rates
and the SSM predictions are
\cite{hahn93}
\begin{equation}
\begin{array}{lcl}
	R_{\rm Ga}\mbox{(GALLEX)} & = & 0.66\pm 0.10
		\vphantom{\Bigg |}\\
	R_{\rm Ga}\mbox{(SAGE)} & = & 0.44^{+0.17}_{-0.21}
		\vphantom{\Bigg |}\\
	R_{\rm Cl} & = & 0.26\pm 0.05 \vphantom{\Bigg |}\\
	R_{\rm Kam} & = & 0.47\pm 0.09 \vphantom{\Bigg |}\ .
\end{array}
\label{E:01}
\end{equation}
The deviation of $R$ from unity is called
the Solar neutrino problem or puzzle.

Another important source of non-accelerator neutrinos is the cosmic rays.
In addition to $\nu_e$, $\nu_\mu$ and $\overline{\nu_e}$,
$\overline{\nu_\mu}$ from $\pi^\pm$ and
$K^\pm$ decay,
the decay-product muons can also produce muon and electron neutrinos within
the atmosphere and the resulting neutrinos are detected
after traveling through the atmosphere or the Earth.
They are the atmospheric neutrinos.
The ratio of the $\nu_\mu$ (and $\overline{\nu_\mu}$) flux to
the $\nu_e$ (and $\overline{\nu_e}$) flux is roughly two
which has been confirmed by detailed Monte Carlo calculations
\cite{barrgaisserstanev89}
for low energy atmospheric neutrinos
($0.1\mbox{GeV}\lsim E_\nu\lsim 2\mbox{GeV}$).
Two large underground water Cherenkov detectors,
the IMB \cite{becker-szendy92} and
Kamiokande \cite{hirata92},
found the ratios, for the contained events,
	\begin{equation}
	R\left( \frac{\displaystyle \mu}{\displaystyle e} \right)
	=\frac{\displaystyle
		R\left(
		\frac{\displaystyle \mu}{\displaystyle e}
		\right)_{\mbox{\small exp}}
	      }{\displaystyle
		R\left(
		\frac{\displaystyle \mu}{\displaystyle e}
		\right)_{\mbox{\small MC}}
 	      }\ =\
	\left\{ \begin{array}{l}
	\vphantom{\Bigg|}
	0.60^{\displaystyle +0.07}_{\displaystyle -0.06}
	\pm 0.05\hskip 1.0cm \mbox{Kamiokande}\\
	\vphantom{\Bigg|}
	0.54\pm 0.05 \pm 0.12\hskip 1cm \mbox{IMB}\ ,
	\end{array}\right.
	\label{E:001}
	\end{equation}
where $R\left(\mu/e\right)_{\mbox{\small exp}}$ and
$R\left(\mu/e\right)_{\mbox{\small MC}}$ are the observed and Monte Carlo
simulated ratios, respectively,
for the muon and electron events induced in the detectors
by $\nu_\mu$ ($\overline{\nu_\mu}$) and $\nu_e$ ($\overline{\nu_e}$).
The preliminary data from SOUDAN-2 \cite{roback92} also supports
the above results
with $R(\mu/e)=0.55\pm 0.27\pm 0.10$,
whereas the previous Frejus \cite{berger90}and NUSEX \cite{nusex}
experiments failed to see the deviation
of $R\left(\mu/e\right)_{\mbox{\small exp}}$ from unity.
We wish to mention that
this anomaly, in particular the Kamioka data, has about
the same statistical and systematic
significance as in the case of the Solar neutrino
deficit, and, furthermore, is less model dependent
than the Solar neutrino case.
(However, it has been argued by some that unlike the Solar neutrinos, the
electron and muon identification for the atmospheric neutrino experiments
needs more improvement.)

If we take both anomalies seriously,
one of the popular and plausible solutions
is the neutrino oscillation.

In the case of the Solar neutrinos, the matter-enhanced neutrino
oscillations (the Mikheyev-Smirnov-Wolfenstein (MSW) effect
\cite{MikheyevSmirnovWolfenstein}) can explain the problem,
yielding two
distinct allowed regions in the $\Delta m^2$-$\sin^2 2 \theta$ parameter space.
One of the solutions is the so-called large $\sin^2(2\theta)$ solution
with $ \Delta m^2 \sim 10^{-5} \mbox{eV}^2$.
The validity of this solution has been questioned in the past based on the poor
$\chi^2$ fit.
However, this criticism is unwarranted because analyses using the three
generations of neutrinos enlarge the allowed area of the large angle solution
considerably \cite{bottinoinghamkim89}.

The explanation
\cite{KrastevPetcov92,AckerLearnedPakvasaWeiler93,babushafi92}
based on the long-wavelength vacuum oscillations,
on the other hand, is still viable and
leads to the allowed region of the $\Delta m^2-\sin^2 2\theta$ plot
to be around $ \Delta m^2 \simeq 10^{-10} (\mbox{eV})^2$ and
$ 0.75\lsim\sin^2 2\theta\lsim1.0 $.

The most intriguing explanation of the atmospheric neutrino problem is
also due to neutrino oscillations in vacuum, {\it i.e.},
the $\nu_\mu$ ($\overline{\nu_\mu}$) is converted into $\nu_\tau$
($\overline{\nu_\tau}$) depleting its flux whereas
the $\nu_e$ ($\overline{\nu_e}$) flux remains unchanged.
This explanation again yields a solution with a large
	$\sin^2(2\theta)$
value in the
	$\Delta m^2-\sin^2(2\theta)$
plot with
	$10^{-3}\mbox{eV}^2 \lsim\Delta m^2 \lsim 10^{-1}\mbox{eV}^2$.

We note that the common feature of these neutrino oscillation solutions
is the large mixing between two neutrinos and
the results remain valid
as long as the third generation
neutrino is magically decoupled while the other two neutrinos are almost
maximally mixed.
In general, however, we have to include all three neutrino
generations with arbitrary mixing and mass parameters for
fully consistent analysis.

In this paper, however, in order to make the analysis simple,
we assume that the three generation
neutrinos are maximally mixed and keep only the mass parameters free.
The maximal mixing provides us with great mathematical simplicity.
However, this does not imply the loss of generality because as long as
mixing angles are reasonably large, the qualitative results still remain
the same.
The three generation neutrino oscillation with maximal mixing has been
discussed in the past \cite{AckerLearnedPakvasaWeiler93},
but the authors did not take into account of the energy dependence of
the oscillating term by
restricting the mass squared differences to be greater than
 	$10^{-10}(\mbox{eV})^2$.
In this case the survival probability for the solar neutrino,
	$P(\nu_e \to \nu_e)$,
at all energies, is 1/3.
The atmospheric neutrino anomaly has also been
addressed by them by assuming that all
	$\Delta m_{ij}^2$
to be in the range
	$(0.5 - 1.2) \times 10^{-2} \mbox{eV}^2$.

Here, we do not make any assumption on the mass parameters and keep
the energy dependence of the survival probability when we calculate the
event rates for the Solar neutrino experiments and
flux ratios for the atmospheric neutrino experiments.
We demonstrate that, as hinted in the two generation analysis,
as long as
	$\Delta m^2_{21} \sim 10^{-10} \mbox{eV}^2$
and
	$\Delta m^2_{32}=(10^{-3}\sim 10^{-1})\mbox{eV}^2$,
the maximally mixed three generations of neutrino scheme is consistent with
all the data for the Solar and atmospheric neutrinos.
Finally, we show that large neutrino mixing angles with
completely non-degenerate neutrino mass hierarchy can be realized in the
frame work of the see-saw mechanism.

\section{Maximal Mixing among Three Neutrinos}
The oscillation probability for $\nu_\alpha\to\nu_\beta$
($\alpha,\beta$ are flavor indices) for the three generation case is given by
\cite{KimPevsner93}
\begin{eqnarray}
P(\nu_\alpha\to\nu_\beta)&=&(a+b-c)\mid U_{\alpha 1}^\ast\mid^2
					\mid U_{\beta 1}\mid^2
                            +(a-b+c)\mid U_{\alpha 2}^\ast\mid^2
					\mid U_{\beta 2}\mid^2 \nonumber\\
                         &&   +(-a+b+c)\mid U_{\alpha 3}^\ast\mid^2
					\mid U_{\beta 3}\mid^2\; ,
\label{E:alpha_to_beta}
\end{eqnarray}
where $U_{\alpha i}$ and $U_{\beta i}$ are the matrix elements of a $3\times 3$
unitary matrix which relates the weak and mass eigenstates of neutrinos.
In Eq. (\ref{E:alpha_to_beta})
\begin{eqnarray}
	a & \equiv & 2 \sin^2\left(\frac{1.27 L \Delta m_{21}^2}{E}\right)
	\nonumber\\
	b & \equiv & 2 \sin^2\left(\frac{1.27 L \Delta m_{31}^2}{E}\right)
	\label{abc} \\
	c & \equiv & 2 \sin^2\left(\frac{1.27 L \Delta m_{32}^2}{E}\right)\ ,
	\nonumber
\end{eqnarray}
where  $L,E$ and $\Delta
m_{ij}^2\equiv m_i^2-m_j^2$ are in units of meter, MeV and $\mbox{eV}^2$,
respectively.

We are interested in the case where three neutrinos are maximally mixed,
{\it {\it i.e.}}, the mixing matrix $U$ is given by
\cite{nussinov76,wolfenstein78}:
\begin{equation}
	U\; =\; \frac{1}{\sqrt{3}}
		\left( \begin{array}{ccccccc}
			\phantom{a}1\phantom{a} & \phantom{a}-1
			\phantom{a}&\phantom{a} -1
			\phantom{a}\vphantom{\Bigg|}\\
			1 & -x & -x^2 \vphantom{\Bigg|}\\
			1 & -x^2 & -x \vphantom{\Bigg|}
			\end{array}\right)
	\; , \; x\ =\ e^{2\pi i/3}\ .
\label{E:Wolfenstein}
\end{equation}
Neutrino masses are left unconstrained in the model under discussion.
As can be seen in Eq. (\ref{E:Wolfenstein}),
$CP$ must be violated; otherwise $U$ is not unitary.
By substituting Eq. (\ref{E:Wolfenstein})
into Eq. (\ref{E:alpha_to_beta}), we find
\begin{eqnarray}
 	P(\nu_e \to \nu_\mu) & = &
 	P(\nu_\mu \to \nu_\tau)\; = \;
 	P(\nu_e \to \nu_\tau)\;  \nonumber \\
	&=&\frac{2}{9}
	\left(\sin^2 k_{21}+\sin^2 k_{31}+\sin^2 k_{32}	\right),
\label{E:threemaxprob2}
\end{eqnarray}
where
\begin{equation}
	k_{ij}\equiv \frac{1.27L\Delta m_{ij}^2}{E} \ .
\label{E:kij}
\end{equation}
As implied by maximal mixing, the three generations of neutrinos
oscillate into each other with equal probability.
It should be noted that as long as neutrino masses are sufficiently small,
the maximal mixing is phenomenologically consistent with all the previously
known constraints.
\subsection{Solar Neutrinos}
For the solar neutrinos the distance involved is given by
\begin{equation}
L=L_0\left[1-\epsilon \cos\left(2\pi\frac{t}{T}\right)\right],
\label{E:AU}
\end{equation}
where $L_0$ is one Astronomical Unit ($=1.5\times 10^{11}\mbox{m}$),
$\epsilon=0.0167$ is the eccentricity of the Earth orbit and $t$
takes $t=0$ on June 21
($T=365$ days).
Since we are anticipating (see below)
	$\Delta m_{31}^2\simeq \Delta m_{32}^2 \gg 10^{-10} \mbox{eV}^2$
in order to solve the atmospheric neutrino problem,
we have, from Eq. (\ref{E:threemaxprob2}),
\begin{equation}
 	P(\nu_e \to \nu_\mu) \; = \;
 	P(\nu_e \to \nu_\tau)\; = \;
	\frac{2}{9}
	\left(\sin^2 k_{21}+\frac{1}{2}+\frac{1}{2}\right),
\label{E:etomu}
\end{equation}
or
\begin{equation}
 	P(\nu_e \to \nu_e) \; = \;
 	1-\frac{4}{9}
	\left(1+\sin^2 k_{21}\right)\; =\;\frac{5}{9} -\frac{4}{9}
	\sin^2 k_{21}\ .
\label{E:etoe}
\end{equation}
Note that the maximal value of the survival probability of $\nu_e$
is $\left(5/9\right)=0.56$.
This implies that, since the Standard Solar Model predicts the event
rate for the GALLEX
detector to be $132^{+9}_{-6}$ SNU, the observed event rate must be
\begin{equation}
 	\Sigma_{\mbox{\small expt}} \lsim 80\ .
\label{E:gap}
\end{equation}
If the $\Sigma_{\mbox{\small expt}}$
turns out to be greater than the above, this maximal mixing model
(as an explanation with vacuum oscillation) is ruled out (or
	$\Delta m_{32}^2$
and
	$\Delta m_{31}^2$
must take values such that
	$\sin^2k_{32}$
and
	$\sin^2k_{31}$
cannot be approximated by 1/2).
The latest GALLEX rate is
$\Sigma_{\mbox{\small expt}} = 87\pm14\pm7$
which is still consistent with Eq. (\ref{E:gap}).
By using the formula for the observed event rate

\begin{equation}
	\Sigma=\sum_i\int\sigma_i(E_\nu)\phi_i(E_\nu)
			 P(\nu_e\to\nu_e;E_\nu)dE_\nu,
\label{E:sigma}
\end{equation}
where the summation is for all possible neutrino-production reactions such
as $pp$, ${}^7$Be, ${}^8$B, $\cdots$ processes,
$\sigma_i(E)$ are the detection cross sections and $\phi_i$ are the
SSM neutrino fluxes,
we have carried out numerical calculation of Eq. (\ref{E:sigma}).
The results of the calculations for the Kamioka, Cl and Ga experiments
are shown in Fig. (\ref{F:001}).
\begin{figure}[tb]
\begin{picture}(400,500)(0,0)
\put(50,0){\psfig{figure=multi.ps,height=7in}}
\end{picture}
\begin{center}
{\Large Figure 1}
\end{center}
\label{F:001}
\end{figure}
In the figures, the horizontal solid lines are the central values of the
experiments with the dashed lines representing one standard deviation.
Oscillating solid lines represent the theoretical values obtained from
Eq. (\ref{E:sigma}), again with the dotted lines indicating one
standard deviation resulting from the errors in the SSM.
We can see that in the region of $\Delta m_{21}^2$
\begin{equation}
	4\times 10^{-11}\mbox{eV}^2 \lsim
	\Delta m_{21}^2 \lsim
	2\times 10^{-10}\mbox{eV}^2\ ,
\label{E:m12}
\end{equation}
theory and experiments agree within one standard deviation.
This conclusion is in agreement with that of the two generation analysis.

\subsection{Atmospheric Neutrinos}
\noindent
Let $N_e$ and $N_\mu$ be the original $\nu_e$ and $\nu_\mu$ fluxes,
respectively, at the point of production somewhere in the atmosphere.
Now, due to oscillations after travelling a distance $L$,
the effective flux at the point of detection is
\begin{eqnarray}
N_e^{ef\hskip-0.1cmf} & = & 	  N_e+N_\mu P(\nu_\mu\to \nu_e)
			- N_e P(\nu_e\to \nu_\mu)
			- N_e P(\nu_e\to \nu_\tau)
			\vphantom{\Bigg|} \nonumber\\
			& = &
	N_e \left[ 1- P(\nu_e\to \nu_\mu)
			- P(\nu_e \to \nu_\tau)
			+\left(
		\frac{\displaystyle N_\mu}{\displaystyle N_e}\right)
		P(\nu_\mu\to \nu_e)
	     \right]\ ,
\label{E:efffluxe}
\end{eqnarray}
where $(N_\mu/N_e) \simeq 2.08$
is the calculated ratio of $\nu_\mu$ and $\nu_e$
for low energy neutrinos \cite{barrgaisserstanev89}.

Since we have $L \lsim 13\times 10^3\ \mbox{Km}$,
$E_\nu \gsim 0.2\ \mbox{GeV}$ and
$\Delta m_{21}^2 \sim 10^{-10}\ \mbox{eV}^2$, $\sin^2 k_{21}$ is well
approximated by zero and, because of the assumed hierarchy $m_3>m_2>m_1$,
we expect
\begin{equation}
	\sin^2 k_{32}\simeq\sin^2 k_{31}\ ,
\label{E:13}
\end{equation}
so that
\begin{eqnarray}
	P(\nu_e\to\nu_\mu)&=&P(\nu_e\to\nu_\tau)={\displaystyle 2
	\over\displaystyle 9}\left(\sin^2 k_{21}+\sin^2k_{32}
	+\sin^2k_{31}\right)\nonumber \\
	&\simeq& {\displaystyle 4\over\displaystyle 9}\sin^2k_{32}\equiv P\ .
\label{E:14}
\end{eqnarray}
Therefore, the ratio of the effective and original fluxes is given by
\begin{equation}
	R(\nu_e\to\nu_e)\equiv
	\left(\frac{\displaystyle N_e^{ef\hskip-0.1cmf}}
	      	   {\displaystyle N_e}
	\right)
	=1-2P+2.08P=1+0.08P\simeq 1\  ,
\label{E:15}
\end{equation}
implying that the maximal mixing oscillations practically do not
modify the $\nu_e$ flux.
It is interesting that in our maximal mixing scheme, the electron
neutrino flux naturally remains unchanged
in spite of equal oscillation probabilities among $\nu_e$, $\nu_\mu$
and $\nu_\tau$.

In the case of $\nu_\mu$, we have, using the same notation as above,
\begin{eqnarray}
N_\mu^{ef\hskip-0.1cmf} & = & 	  N_\mu+N_e P(\nu_e\to \nu_\mu)
			- N_\mu P(\nu_\mu\to \nu_e)
			- N_\mu P(\nu_\mu\to \nu_\tau)
			\vphantom{\Bigg|}\nonumber \\
			& = &
	N_\mu \left[ 1- 2\ P	+0.48\ P\right]
	     \; =\; N_\mu \left(1-1.52\ P\right) ,
			\vphantom{\Bigg|}
\label{E:16}
\end{eqnarray}
leading to
\begin{equation}
R\left(\nu_\mu\to\nu_\mu\right)=
	\left(\frac{\displaystyle N_\mu^{ef\hskip-0.1cm f}}
	           {\displaystyle N_\mu}
\right)\; =\; 1-1.52 \ P\; .
\label{E:17}
\end{equation}
Since we have $P\leq 4/9$ [see Eq. (\ref{E:etoe})],
the maximal mixing oscillations can deplete $\nu_\mu$ by as much as
80\%.
{}From Eqs. (\ref{E:15}) and (\ref{E:17})
the effective $\nu_\mu$ and $\nu_e$ flux ratio is given by
\begin{equation}
\frac{\displaystyle R(\nu_\mu\to\nu_\mu)}
     {\displaystyle R(\nu_e\to\nu_e)}
=
\frac{\displaystyle 1-1.52 P}
     {\displaystyle 1+0.08 P}
\simeq 1-1.6 P\; .
\label{E:18}
\end{equation}
The observed $\mu/e$ ratio at the Kamiokande
in which most of theoretical uncertainties in the
Monte Carlo calculations are cancelled out is 0.60
for the ``contained events''. This implies, in this naive
estimate, that, by setting the right-hand side of Eq. (\ref{E:18})
equal to 0.60,
	\begin{equation}
	P(\nu_\mu\to\nu_\tau)=\frac{\displaystyle 4}{\displaystyle 9}
	\sin^2\left(1.27 \Delta m_{32}^2\left\langle
	\frac{\displaystyle L}{\displaystyle E_\nu}\right\rangle\right)
	\simeq 0.25\; .
	\label{E:19}
	\end{equation}
or
	\begin{displaymath}
	\sin^2\left(1.27 \Delta m_{32}^2\left\langle
	\frac{\displaystyle L}{\displaystyle E_\nu}\right\rangle\right)
	\simeq 0.56\; .
	\end{displaymath}
It is interesting to note that the required value of
	$\sin^2\left(1.27 \Delta m_{32}^2\left\langle
	\frac{L}{E_\nu}\right\rangle\right)$
to explain the data is very close to 1/2.
This can be realized in two ways.
First, the oscillation is already in the rapid oscillation region, leading
to
	$\sin^2\left(1.27 \Delta m_{32}^2\left\langle
	\frac{L}{E_\nu}\right\rangle\right)\simeq 1/2$
which implies
	$1.27 \Delta m_{32}^2\left\langle \frac{L}{E_\nu}\right\rangle$
is much larger than unity.
Since
	$\left\langle \frac{L}{E_\nu}\right\rangle \gsim 10^2
	(\frac{\mbox{\small Km}}{\mbox{\small GeV}})$,
we have
	$\Delta m_{32}^2 \gsim  10^{-2} \mbox{eV}^2$.
Another possibility is that
	$\sin^2\left(1.27 \Delta m_{32}^2\left\langle
	\frac{L}{E_\nu}\right\rangle\right)$
happens to be just 0.56.
Since the average value of
	$\langle L/E_\nu \rangle$
for the fully
contained events is estimated to be
	$10^2(\frac{\mbox{\small Km}}{\mbox{\small GeV}})\lsim$
	$\langle \frac{L}{E_\nu} \rangle\ $
	$\lsim 10^3(\frac{\mbox{\small Km}}{\mbox{\small GeV}})$,
we find, in this case,
	\begin{equation}
		10^{-3} \mbox{eV}^2\lsim
		\Delta m_{32}^2 \lsim  10^{-2} \mbox{eV}^2\ .
	\label{E:20}
	\end{equation}
Combining both possibilities (which we cannot distinguish at present)
and the fact that
	$\Delta m_{32}^2 \gsim (10^{-1}\sim 1) \mbox{eV}^2$
is already eliminated for the large mixing case from laboratory oscillation
experiments, we can conclude that
	\begin{equation}
		10^{-3} \mbox{eV}^2\lsim
		\Delta m_{32}^2 \lsim (10^{-1}\sim 1) \mbox{eV}^2\ .
	\label{E:201}
	\end{equation}

The calculated flux for the upward through-going muons is subject to
uncertainties in the cosmic ray flux.
It has been shown in Ref. \cite{FratiGaisserMannStanev93}
that the ratio of the observed number
of such muons and the calculated values ranges from 0.84 to 0.94.
If we interpret the deviation of these ratios from unity as an indication
of the $\nu_\mu$ depletion due to oscillations,
we have, from Eq. (\ref{E:17}),
	\begin{equation}
	(0.84\sim 0.94)=1-1.52\cdot\frac{\displaystyle 4}{\displaystyle 9}
	\sin^2\left(1.27 \Delta m_{32}^2
	\times 10^2\right)\ ,
	\label{E:21}
	\end{equation}
where we have used the value
	$\langle L/E_\nu \rangle\ \simeq 10^2\ (\mbox{Km}/\mbox{GeV})$
\cite{FratiGaisserMannStanev93} and
	$\Delta m_{32}^2$
is in units of
	$\mbox{eV}^2$.
Equation (\ref{E:21}) yields, with the constraint given by Eq. (\ref{E:201}),
	\begin{equation}
	\Delta m_{32}^2=(2.47\ n+0.2\sim 2.47\ n+0.4)
			\times 10^{-2}\ \mbox{eV}^2\; ,
	\label{E:22}
	\end{equation}
where $n=0,1,2,\cdots$.

In contrast to the case we have just discussed, the ratio of the upward
stopping muons and the upward through-going muons is practically free
of the flux normalization
uncertainties.
The calculated ratio for the Kamiokande detector ranges from 0.28 to 0.30
\cite{FratiGaisserMannStanev93},
indicating the insensitivity to the method of calculations.
Unfortunately, the relevant experimental data are not yet available
from Kamiokande.
A preliminary analysis of the IMB-3 group has not come up with any obvious
discrepancy between the observed and calculated ratios.

If our interpretation of $\nu_\mu$ oscillations is valid,
we expect the observed ratio to be different from the above
calculated ratio by a factor
	\begin{equation}
	R\left(\frac{\displaystyle S}{\displaystyle T}\right)
	=\frac{\displaystyle 1-1.52
	P\left(\left\langle \frac{\displaystyle L}{\displaystyle E_\nu}
	\right\rangle
	=10^3\left(\frac{\displaystyle \mbox{Km}}{\displaystyle \mbox{GeV}}
	\right)\right)}
	{\displaystyle 1-1.52
	P\left(\left\langle \frac{\displaystyle L}{\displaystyle E_\nu}
	\right\rangle
	=10^2\left(\frac{\displaystyle \mbox{Km}}{\displaystyle \mbox{GeV}}
	\right)\right)
	}\; ,
	\label{E:25}
	\end{equation}
since the average values of $\langle L/E_\nu \rangle $ for the neutrinos
which produce the upward stopping muons and the upward through-going muons
are given by
	$\sim 10^3\ (\mbox{Km}/\mbox{GeV})$
and
	$\sim 10^2\ (\mbox{Km}/\mbox{GeV})$,
respectively.
If the oscillation is already in the rapid oscillation region ({\it i.e.},
	$\Delta m_{32}^2\gsim 10^{-2}\mbox{eV}^2$
) the above ratio is, of course, unity.
However, if that is not the case, the ratio can be different from unity.
For example, the above factor takes the following values,
	\begin{equation}
	R\left(\frac{\displaystyle S}{\displaystyle T}\right)=
	\left\{
	\begin{array}{lcl}
	0.38 &\mbox{for}\vphantom{\Big |} & \Delta m_{32}^2=
	10^{-3}\mbox{eV}^2\\
	0.82 &\mbox{for}\vphantom{\Big |} & \Delta m_{32}^2=
	3\times 10^{-3}\mbox{eV}^2
	\end{array}
	\right. \ ,
	\label{E:27}
	\end{equation}
which can soon be tested in the future.

\section{See-Saw Enhancement of Neutrino Mixing}

So far we have conjectured that three neutrinos are maximally mixed,
motivated by the vacuum oscillation solutions for the solar neutrino and
atmospheric neutrino problems.
The simultaneous solution requires the neutrino mass hierarchy ,
	$m_1  \ll m_2  \ll m_3 $
and the maximal mixings.

The obvious question then is ``Is it possible for completely non-degenerate
(in mass) neutrinos to have maximal, or at least very large mixings ?''
The conventional wisdom tells us that mixing angles are determined by
lepton mass ratios.
In the quark sector, the Cabibbo angle is well reproduced by the formula
	$\tan^2\theta_{_C}=m_d/m_s$.
In the same spirit, the neutrino mixing angle, in the case of two generations,
is expected to be determined by the ratios
	$m_e/m_\mu$ and $m(\nu_1)/m(\nu_2)$.
If
	$m(\nu_1) \ll m(\nu_2)$,
one expects the neutrino mixing angle
to be very small.

However, this conclusion is not valid when one invokes the see-saw mechanism.
The possibility that the see-saw mechanism may enhance lepton mixing up
to maximal was first discussed by Smirnov\cite{Smirnov93}.
When the see-saw mechanism is invoked with more than one generation of
neutrinos, mixing angles
for the resultant light Majorana neutrinos become dependent
upon masses of heavy right-handed Majorana neutrinos which are usually
provided by GUTS models.
In particular, when certain relationships among the mass parameters
in the original Dirac and heavy right-handed Majorana mass matrices
are satisfied,
mixing angles for the light Majorana neutrinos can substantially be enhanced.
Let us consider the $4\times 4$
mass matrix, ({\it i.e.}, the case of the two generations)
\begin{equation}
	{\cal M}=\left( 	\begin{array}{cc}
			\phantom{a}\mbox{{\bf 0}}\phantom{a} &
			\phantom{a}\mbox{{\bf m}}\phantom{a}
			\vphantom{\Bigg|}\\
			\phantom{a}\mbox{{\bf m}}\phantom{a} &
			\phantom{a}\mbox{{\bf M}}\phantom{a}
			\vphantom{\Bigg|}
			\end{array}
	   \right)\ ,
\label{E:1_1}
\end{equation}
where {\bf m} and {\bf M} are $2\times 2$ real and symmetric matrices,
representing the Dirac and heavy Majorana neutrino mass matrices,
respectively.

Let us define $U(\theta_{_D})$ and $U(\theta_{_M})$ such that
\begin{eqnarray}
	U^T(\theta_{_D})\ \mbox{{\bf m}}\
	U(\theta_{_D})&=&\hat{\mbox{{\bf m}}}\ ,
	\nonumber \\
	U^T(\theta_{_M})\ \mbox{{\bf M}} \
	U(\theta_{_M})&=&\hat{\mbox{{\bf M}}}\ ,
\label{E:1_2}
\end{eqnarray}
where $\hat{\mbox{{\bf m}}}$ and $\hat{\mbox{{\bf M}}}$ are
$2\times 2$ diagonal matrices.
By block-diagonalizing the matrix $\cal M$, we find, assuming
$\mid m_{ij} \mid \ll \mbox{det{\bf M}}$ in
the spirit of the see-saw mechanism, the
Majorana mass matrix for light neutrinos which are completely
decoupled from super heavy ones as

\begin{equation}
	\mbox{{\bf m}}_{\mbox{\tiny Maj}}=\mbox{{\bf m}}\  \mbox{{\bf M}}^{-1}
	\ \mbox{{\bf m}}\ .
\label{E:1_3}
\end{equation}
Solving Eq. (\ref{E:1_2}) for {\bf m} and {\bf M} and substituting them into
Eq. (\ref{E:1_3}), we obtain
\begin{eqnarray}
	\mbox{{\bf m}}_{\mbox{\tiny Maj}} &=&
	\left[ U(\theta_{_D})\hat{\mbox{{\bf m}}} U^T(\theta_{_D})\right]
	\left[ U(\theta_{_M})\hat{\mbox{{\bf M}}}^{-1} U^T(\theta_{_M})\right]
	\left[ U(\theta_{_D})\hat{\mbox{{\bf m}}} U^T(\theta_{_D})\right]
	\nonumber \\
	&\equiv & U(\theta_{_D})\mbox{{\bf m}}_{ss} U^T(\theta_{_D})\ ,
\label{E:1_4}
\end{eqnarray}
where
\begin{equation}
	\mbox{{\bf m}}_{ss} \equiv
	 \hat{\mbox{{\bf m}}}U (\theta_{_M}-\theta_{_D})
	\hat{\mbox{{\bf M}}}^{-1}
	U^T(\theta_{_M}-\theta_{_D})\hat{\mbox{{\bf m}}}\ ,
\label{E:1_5}
\end{equation}
and we have used $U^T(\alpha)U(\beta)=U^T(\alpha-\beta)$
and $U^T(\beta-\alpha)=U(\alpha-\beta)$.
Defining the orthogonal matrix that diagonalizes {\bf m}$_{ss}$ in
Eq. (\ref{E:1_5}) as $U(\theta_{ss})$, we can rewrite Eq. (\ref{E:1_4}) as
\begin{equation}
	\mbox{{\bf m}}_{\mbox{\tiny Maj}} =
	 U(\theta_{_D}+\theta_{ss})\hat{\mbox{{\bf m}}}_{ss}
	 U^T(\theta_{_D}+\theta_{ss})\ ,
\label{E:1_6}
\end{equation}
implying that the overall mixing angle $\theta_\nu$ is now
\begin{equation}
	\theta_\nu = \theta_{_D} + \theta_{ss}\ .
\label{E:1_7}
\end{equation}
Therefore, even if the angle $\theta_{_D}$ is originally small,
the neutrino mixing angle can be large if the angle due to the
see-saw mechanism is large.

We now explicitly demonstrate the possibility of a large
$\theta_\nu = \theta_{_D} + \theta_{ss}$.
In order to illustrate the case in point
in a more transparent way than originally presented by Smirnov,
let us consider following mass matrix without imposing any restrictions on
{\bf m} and {\bf M}.
\begin{equation}
	{\cal M} =
	\left( \begin{array}{cccc}
		\phantom{a}0\phantom{a} &\phantom{a}0\phantom{a}
	 	 &\phantom{a}m_1\phantom{a} &\phantom{a}m\phantom{a}
		\vphantom{\Bigg|}\\
		0&0&m&m_2\vphantom{\Bigg|}\\
		m_1&m&M_1&M\vphantom{\Bigg|}\\
		m&m_2&M&M_2\vphantom{\Bigg|}\\
		\end{array}
	\right)\begin{array}{l}\vphantom{\Bigg|}\\
	\vphantom{\Bigg|} \\ \vphantom{\Bigg|} \\ .\end{array}
\label{E:1_8}
\end{equation}
We are assuming, in the spirit of the see-saw mechanism,
\begin{equation}
	\mid m\mid , \mid m_i\mid \ll \mbox{det{\bf M}}.
\label{E:1_9}
\end{equation}
After block-diagonalizing the matrix ${\cal M}$, we obtain
\begin{equation}
\begin{array}{l}
\mbox{{\bf m}}_{\mbox{\tiny Maj}}=
	\frac{\displaystyle 1}{\displaystyle M_1 M_2-M^2}\\
\phantom{aaa}\times
	\left( \begin{array}{cc}
		\phantom{a}m_1^2M_2-2m_1mM+m^2M_1\phantom{a}
		&\phantom{a}\begin{array}{l}
				m(m_1M_2+m_2M_1)
				\\ \phantom{aaaaaa}-M(m_1m_2+m^2)
		\end{array}\phantom{a}
		\vphantom{\Bigg|}\\
		\phantom{a}\begin{array}{l}
				m(m_1M_2+m_2M_1)
				\\ \phantom{aaaaaa}-M(m_1m_2+m^2)
		\end{array}\phantom{a}
		&m_2^2M_1-2m_2mM+m^2M_2
		\vphantom{\Bigg|}\\
		\end{array}
	\right)\begin{array}{l}\vphantom{\Bigg|}\\.\end{array}
\end{array}
\label{E:1091}
\end{equation}
It is  straight-forward to calculate
$\hat{\mbox{{\bf m}}}_{ss}\equiv \mbox{diag}(\lambda_1,\lambda_2)$
and $U(\theta_{_D}+\theta_{ss})$. The results are
\begin{eqnarray}
	\lambda_{1,2}&=&
	\frac{\displaystyle  1}
	     {\displaystyle  2(M_1 M_2-M^2)}
	\left\{	\begin{array}{l}
			m_1^2 M_2-2 m m_1 M + m^2 M_1\\
			\phantom{Bigg ld }+m_2^2 M_1-2m m_2 M + m^2 M_2
		\end{array}
      \vphantom{\left[\begin{array}{l}M^2\\M^2\end{array}\right]^{\frac{1}{2}}}
	\right. \label{E:1_10}\\
	&& \left.
	\phantom{21}
       	\pm
	\left[	\begin{array}{l}
			[(m_1^2 M_2-2 m m_1 M + m^2 M_1)
			- (m_2^2 M_1-2m m_2 M + m^2 M_2)]^2\\
			\phantom{Biace shouldaaaaa}
			+
			4 [ m ( m_1 M_2+m_2 M_1) - M (m_1 m_2 + m^2 )]^2
		\end{array}
	\right]^{\frac{1}{2}} \right\}\ ,
	\nonumber
	\end{eqnarray}
and
	\begin{equation}
	\tan\left[2(\theta_{_D}+\theta_{ss})\right] =
	\frac{\displaystyle \vphantom{\Bigg|}
	2 \left[ m ( m_1 M_2+m_2 M_1) - M (m_1 m_2 + m^2 )\right]}
	{\displaystyle \vphantom{\Bigg|}
	(m_2^2 M_1-2m m_2 M + m^2 M_2)
	-(m_1^2 M_2-2 m m_1 M + m^2 M_1)}\ .
	\label{E:1_11}
	\end{equation}

The most obvious condition for the maximal mixing and mass
hierarchy is that the four elements of the matrix
	$\mbox{{\bf m}}_{\mbox{\tiny Maj}}$
be the same. In other words,
the matrix elements of {\bf m} and {\bf M} should satisfy
	\begin{equation}
 	(m_2^2 - m^2) \left(\frac{\displaystyle M_1}{\displaystyle M_2}\right)
	+2 m( m_1 - m_2)
	\left(\frac{\displaystyle M}{\displaystyle M_2}\right)
	  = m_1^2 - m^2\ ,
	\label{E:10111}
	\end{equation}
and
\begin{equation}
 	(m_2 - m)^2 \left(\frac{\displaystyle M_1}{\displaystyle M_2}\right)
	+2 ( m_1 - m)( m_2 - m)
		\left(\frac{\displaystyle M}{\displaystyle M_2}\right)
	  = -(m_1 - m)^2\ .
\label{E:10112}
\end{equation}
where $M_2$ is assumed to be finite.
Eq. (\ref{E:10111}) comes from matching two diagonal elements of
the mass matrix in Eq. (\ref{E:1091}) and Eq. (\ref{E:10112}) is obtained
by equating the sum of
diagonal elements and that of off-diagonal elements.
Now we can solve the equations for $M_1/M_2$ and $M/M_2$.
We find, from Eqs. (\ref{E:10111}) and (\ref{E:10112}),
\begin{equation}
	\frac{\displaystyle M_1}{\displaystyle M_2}=
		\frac{\displaystyle (m- m_1)^2}{\displaystyle (m-m_2)^2}
	\phantom{aa},\phantom{aa}
	\frac{\displaystyle M}{\displaystyle M_2}=
		-\frac{\displaystyle m-m_1}{\displaystyle m-m_2}\ ,
\label{E:102}
\end{equation}
for which, the mixing angle $(\theta_{_D}+\theta_{ss})$ becomes $45^0$,
and $\lambda_1=0$ and $\lambda_2$ is finite, as desired.
It is clear, however, that in this case the determinant of {\bf M} vanishes
so that the inverse of the matrix does not exist.
To remedy this problem, we have to relax one or both conditions
of Eqs. (\ref{E:10111}) and (\ref{E:10112}).
Since the condition of Eq. (\ref{E:10111}) is directly related
to the maximal mixing, one is forced to relax Eq. (\ref{E:10112}).

Now, with one less condition to satisfy, we will give several
examples in which the maximal mixing is realized with significant
hierarchy in the mass eigenvalues.

\noindent {\bf Case 1.}   The first case is for $M=0$.
The obvious possibility in this case is that all the elements of {\bf m} are
identical. The mass eigenvalues of light Majorana neutrinos
are $(0,2m^2(M_1+M_2)/M_1M_2)$ and they have extreme hierarchy,
{\it i.e.}, one of
the eigenvalues is zero and the other finite.
In this case, the see-saw mechanism provides only the smallness of the
light Majorana neutrinos while the maximal mixing with the mass hierarchy
is the result of the nature of the matrix {\bf m}.

\noindent {\bf Case 2.}
The next possibility is when $M_1$ and $M_2$ are adjusted
with still keeping $M=0$ so that
the condition $\tan\left[2(\theta_{_D}+\theta_{ss})\right]=\infty$
is met, {\it {\it i.e.}},
$$
	\frac{\displaystyle M_1}{\displaystyle M_2}
	=
	\frac{\displaystyle m^2-m_1^2}{\displaystyle m^2-m_2^2}\ .
$$
Then, the ratio of two eigenvalues is given by
\begin{equation}
	\frac{\displaystyle \lambda_1}{\displaystyle \lambda_2}
	=
	\frac{\displaystyle (m_1-m)(m_2-m)}{\displaystyle (m_1+m)(m_2+m)}\ .
\label{E:103}
\end{equation}
When $m_1\simeq m$ and/or $m_2\simeq m$,
$\lambda_1/\lambda_2$ can be very small and  $M_1$ and $M_2$ can have a very
wide range of ratios.

\noindent {\bf Case 3.}   Another case of interest
is when the neutrino Dirac mass matrix has the Fritzsch form,
{\it {\it i.e.},} $m_1=0$.
Equation (\ref{E:10111}) then reduces to
\begin{equation}
(m_2^2-m^2) M_1-2mm_2M+m^2M_2=0\ ,
\label{E:104}
\end{equation}
and the hierarchical condition reduces to
\begin{equation}
	mM_1\simeq \mid m_2 M_1-mM\mid
\label{E:105}
\end{equation}
Equation (\ref{E:104}) guarantees the maximal mixing and Eq. (\ref{E:105})
implies a desired mass hierarchy.

%
%
\section{Conclusion}
Motivated by recent indications of the possible large mixing angle solutions
for both the Solar and atmospheric neutrino problems, we have
examined the scheme in which all three neutrinos are maximally mixed.
In this model, only the neutrino masses are treated as unknown parameters.
We have analyzed the most recent data on the Solar and atmospheric neutrinos
using this model.
All three types of the Solar neutrino data (Cl, Water, Ga) can be explained
by vacuum oscillations of the maximally mixed three generations of neutrinos
if
	$\Delta m^2_{21}\equiv m^2_2-m^2_1\simeq 10^{-10}\mbox{eV}^2$
and
	$\Delta m^2_{32} \gg 10^{-10}\mbox{eV}^2$,
as suggested
by analysis based on the two
generation treatment.
That is, addition of the third generation of neutrino does not change the
prediction of the two generation analysis.
One definite prediction of the three generation model
that the observed event rate for the
Ga experiment cannot be larger than 80 SNU can soon be tested as the
Ga data improves in the near future.

Similarly, the current data on the atmospheric neutrino experiments can
be explained in this model for
	$10^{-3} \mbox{eV}^2
	\lsim \Delta m_{32}^2 \lsim
	(10^{-1}\sim 1) \mbox{eV}^2$
and
	$\Delta m^2_{21} \simeq 10^{-10}\mbox{eV}^2$.
It is interesting to note that in spite of the equal values of
$P(\nu_e\to \nu_\mu)$, $P(\nu_\mu\to \nu_\tau)$ and $P(\nu_e\to \nu_\tau)$ in
our model, the atmospheric $\nu_e (\overline{\nu_e})$ flux naturally remains
the same whereas that of $\nu_\mu (\overline{\nu_\mu})$ is substantially
suppressed as indicated by the data.

We have demonstrated that although the conventional wisdom learned from the
quark sector suggests small mixing angles in the lepton sector,
it is possible to have large or maximal mixings for the neutrinos
if the see-saw mechanism is invoked to generate small neutrino masses.
We have presented explicit two-generation examples (or models) in which
the two light Majorana neutrinos are maximally mixed even though their masses
are highly non-degenerate.

\section*{Acknowledgments}
This work has been supported in part by the National Science Foundation.

\newpage
\section*{Figure Caption}

Fig. 1: Comparison of Calculated and Observed Event rates for Kamioka, Cl
and Ga Experiments.
\end{document}